\documentclass[doublecol]{epl2}

\usepackage{amsmath}
\usepackage{amssymb}
\usepackage{graphicx}
\usepackage{graphics}

\newcommand{\D}{\mathcal{D}}
\newcommand{\Tr}{\mathrm{Tr}}
\newcommand{\tr}{\mathrm{tr}}
\newcommand{\T}{^\mathrm{T}}
\newcommand{\openone}{1}

\title{Non-linear $\sigma$ model study of magnetic dephasing
in a mesoscopic spin glass}

\author{Andrei A. Fedorenko \and David Carpentier}

\institute{ Laboratoire de Physique de l'Ecole Normale Sup{\'e}rieure de Lyon, 46,
All{\'e}e d'Italie, 69007 Lyon, France }


\abstract{
We propose  a nonlinear sigma model for the description
of quantum transport in a mesoscopic metallic conductor with magnetic
impurities frozen in a spin glass phase. It accounts for the presence
of both the corresponding scalar and  magnetic random potentials.
In a spin glass, this magnetic random potential is correlated between
different realizations.  As the strength of the magnetic potential is varied,
this model describes  the crossover between orthogonal and unitary universality
classes of the nonlinear sigma model.  We apply this technique to the calculations
of the  correlation of conductance between two frozen  spin configurations
in terms of dephasing rates for the usual low energy modes of weak localization theory.
}

\begin{document}

\maketitle
\section{Introduction}
In a metallic conductor of $\mu m$ size, the interplay between quantum coherence
of electrons  and disorder is at the origin
of several remarkable phenomena at low temperature \cite{Houches95,Akkermans:2007}.
The interferences between different diffusive paths in
the sample lead to a strong dependance of the conductance on the disorder realization
through the geometry of these diffusive paths. In particular the conductance exhibits
universal sample to sample fluctuations and reproducible random variations as a function
of a transverse magnetic flux.
These coherent transport phenomena are naturally
suppressed by various perturbations which are usually referred to as
dephasing sources.
This includes inelastic scattering, for example from phonons, free magnetic
impurities, other electrons.
In this case two electrons following the same path
encounter different potentials. This leads to a random relative phase between them
and the suppression of interference effects.
Such a dephasing source acts on the electrons
themselves, and is usually taken into account phenomenologically
through a dephasing rate $\gamma_{\phi}$ for the electrons.  Elastic scattering by
symmetry breaking potentials  by {\it e.g}
random spin-orbit interactions or frozen magnetic impurities
is also considered as another source of dephasing.
In this case, the origin of dephasing is physically
different from that in the inelastic case. Quantum corrections to transport result
from interference between  electrons travelling along loops of diffusive paths
either in the same directions (Diffuson modes) or in the opposite directions (Cooperon modes).
In the presence of magnetic disorder, the spins of counter-propagating electrons
experience different sequences of rotations.
As a result the corresponding interferences are gradually suppressed   for longer and
longer loops. This suppression is usually interpreted as dephasing of the corresponding
Cooperon modes.
Moreover the magnetic disorder selects out the Diffuson
modes not affected by this dephasing: the relative phase of two electrons with
spins forming a singlet state is insensitive to this spin rotation sequence,
in contrast to that of triplet states.
Hence, this symmetry-breaking elastic scattering induces a spatial decay of these
particular diffusion modes, effectively accounting for this electron's
dephasing phenomenon. As a particular consequence, the magnitude of
the universal conductance
fluctuations is determined by symmetry properties. 

This magnetic dephasing of electrons and its signature on coherent transport has been
recently proposed as a promising probe of spin glass physics \cite{Carpentier:2008}.
In the spin glass phase, which is a fascinating but poorly understood
state of matter \cite{Vincent:2009}, impurity spins $\{\vec{S}_{i}\}$ coupled by
frustrating interactions freeze below the transition temperature $T_{g}$.
Of particular interest for its understanding  are the correlations between
different configurations of spins $\{\vec{S}^{(1)}_{i}\}$ and $\{\vec{S}^{(2)}_{i}\}$
corresponding to either different times \cite{Fisher:1986, Cugliandolo:2002},
or different quenches below $T_{g}$ .
The correlation of conductance between these two spin configurations
(corresponding to different quenches or times) is determined by the
correlation of magnetic dephasing of the electrons in these two
configurations.  This correlation can be expressed in terms of the
dephasing rates for the diffusons and cooperons
composed of electrons traveling along the same path but experiencing
two different magnetic potentials corresponding to two different
spin glass configurations.
Hence, such measurement of correlations of conductance open the route a direct probe of
correlations between frozen spin configurations \cite{Altshuler:1985, Carpentier:2008}.
This naturally requires a  quantitative description of this relative magnetic dephasing
between different spin configurations,  on which we focus in this Letter.

Beyond the perturbative diagrammatic  theory \cite{Akkermans:2007}, a natural
framework for describing the statistics of conductance is the field theoretical
nonlinear sigma model \cite{Altshuler-book}. This powerful field theoretical method
is an essential tool in describing the weak localization regime when computing higher moments
of the conductance, or incorporating spatial or time correlations of the random potential.
 The aim of the present Letter is to apply this technique
to study the quantum electronic transport in a mesoscopic metallic glass.
 This amounts to incorporate  the presence of both a scalar random potential and a
 weaker magnetic  potential.
We will show that this method allows for a very efficient and
elegant description of the corresponding magnetic dephasing, by treating
all elastic scattering potentials on the same footing.
Altland showed how the inelastic dephasing effects  can be phenomenologically
accounted for within this nonlinear sigma model \cite{Altland92}. While his
approach required the introduction of a new fictitious scalar potential,  in the
present context  the source of dephasing is already an  elastic scattering potential.
Here, this dephasing accounts for  the crossover behavior between
different universality classes which characterize  in particular universal weak localization
properties \cite{Hikami80}.
These classes encode the number and nature
of independent diffusive modes contributing to the quantum correction of conductance
for weak disorder.
In the present case, without magnetic impurities the statistics of conductance is
described by the orthogonal class with degenerate diffusive states.
When adding magnetic disorder, the level degeneracy is lifted and simultaneously the
universality class is restricted to the  unitary class. Studying the magnetic dephasing
induced by frozen magnetic impurities amounts  to study the cross-over between these two
universality classes. This is achieved by considering  all massless modes of the
orthogonal case, and describing the different gap openings with  increasing magnetic
disorder, thereby extending non-perturbatively the work of \cite{Efetov80} to the spin
glass physics of interest here.


\section{The nonlinear sigma model}
We consider a $d$ dimensional mesoscopic metallic sample of size $L$ containing impurities
inducing two different types of random scattering of the conduction electrons:
(i) nonmagnetic random scalar potential $V(r)$ coupled to the local fermionic density
 as $V(r) \bar{\psi}(r)\psi(r)$. This potential
is assumed to be Gaussian  with $\langle V(r) \rangle=0$  and variance
\begin{equation} \label{VV}
\langle V(r)V(r') \rangle =\frac1{2\pi \nu_0 \tau_v}\delta(r-r'),
\end{equation}
where $\nu_0$ is the one-electron density of states and $\tau_v$ - the corresponding
elastic mean free time. (ii) a magnetic disorder $\vec{U}(r)$ originating from a collection
of frozen magnetic impurities $\vec{S}_{i}$ with  coupling to the electron's density of spins
$ \vec{U}(r)\cdot  \bar{\psi}(r)\vec{\sigma}\psi(r) $ with $\vec{\sigma}$ being
the Pauli's matrices. $\vec{U}(r)$ is a three dimensional
field taken Gaussian with zero mean in the spin glass state.
Here we focus on both the conductance fluctuations
in a single magnetic disorder configuration, and conductance correlations between
different magnetic disorder realizations  corresponding to different spin glass states.
For the sake of simplicity we consider  correlations between
different realizations of magnetic potential  indexed by $u$ of the form
\begin{eqnarray} \label{UU}
  \overline{U_{u;i}(r)U_{u';j}(r')}=\frac{q_{uu'}}{3}  \delta_{ij} \frac1{2\pi \nu_0 \tau_s} \delta(r-r'),
\end{eqnarray}
where
$q_{uu'}$ is the standard overlap in spin glass theory \cite{Mezard:1987,Parisi:2007},
$q_{uu} =1$ and $\tau_s$ is the elastic mean free time for scattering
by magnetic impurities. Here we neglect possible correlations between the magnetic and
non-magnetic impurities.

The conductance of the sample at frequency $\omega$ for non-interacting electrons
at zero temperature is given by Kubo formula ($\hbar=m=e=1$):
\begin{multline}
  G(\omega) =\\
   \frac{1}{2\pi L^2 } \Tr\left[ \hat{j}_x \mathcal{G}_{\epsilon_F}^{R}
  (r',s'; r,s)\  \hat{j}_{x'} \mathcal{G}_{\epsilon_F-\omega}^{A} (r,s;r',s') \right], \label{G1}
\end{multline}
where $\mathcal{G}^{R,A}$ are the retarded and advanced Green functions,
$s,s'= \uparrow,\downarrow$ - spin variables, and
$\hat{j}_x = i {\partial}/{\partial x}$ - the probability current operator.

The Green functions $\mathcal{G}^{R,A}$
for electrons in the random potentials (\ref{VV}) and (\ref{UU})
can be expressed with the help of a functional integration over Grassmann conjugated
fields $\psi$ and $\bar{\psi}$ as follows
\begin{eqnarray}
\mathcal{G}^{R,A}_{\epsilon_F\pm\omega^+/2} (r,s;r',s') = \frac{-i}{Z^{R,A}}
\int D \bar{\psi} D \psi\  \bar{\psi}(r) \psi(r') e^{iS_{\pm}}
\end{eqnarray}
with the actions
\begin{eqnarray}
S_{\pm} = \int_r \bar{\psi}_s(r)\left[\epsilon_F
\pm \frac12\omega^{+}-H_{ss'}\right] \psi_{s'}(r),
\end{eqnarray}
where $H_{ss'}$ is the Hamiltonian, $\omega^+=\omega+i\delta$ and $\pm$ correspond
to the retarded and advanced Green functions.
 Following ref.~\cite{Efetov80} we define the covariant
and contravariant bispinors
$   \bar{\eta} =  \left(C \eta \right)\T
= \frac1{\sqrt{2}} \left(
   -{\psi}_{\uparrow},
   -{\psi}_{\downarrow},
   \bar{\psi}_{\downarrow} ,
   -\bar{\psi}_{\uparrow} \right)$,
which are related by the charge-conjugation matrix $C = i \sigma_1 \otimes \sigma_2$.
The four components of the bispinor are the two "charge" degrees and the two spin degrees of
freedom. The auxiliary "charge" degree of freedom
is introduced for taking into account on the same footing the two
possible pairing $\psi \bar{\psi}$ and  $\bar{\psi} \bar{\psi}$ which contribute
to the slow part of the free energy.
In order to generate products of the advanced and retarded Green
functions we double the degrees of freedom introducing the index $p = R,A$.
For our purpose we also introduce $2$ copies of the original system with different
configurations of magnetic disorder enumerated by $u=1,2$.
Then in terms of the bispinors the action for the noninteracting electrons can
be written as
\begin{eqnarray} \label{action1}
  \mathcal{S}=\int d^d{r}\ \bar{\eta}_u(r)\left[
   \hat{\xi}+\frac12 \omega^+\Lambda - V(r) - \vec{U}_u(r)\cdot\vec{\Sigma}
  \right] \eta_u(r),
\end{eqnarray}
where  $\hat{\xi}= \epsilon_F-{\hat{p}^2}/{(2m)}$, $\Lambda=(\sigma_3)_{p,p'}\otimes \openone$
and $\vec{\Sigma} = \sigma_3 \otimes \vec{\sigma}$.

To compute the dimensionless conductance correlation in a given sample with two different
magnetic disorder configurations
$(\Delta G)_{uu'}^2= \overline{\langle  G(V,\vec{U}_u)G(V,\vec{U}_{u'}) \rangle}_c$
we generalize the formalism of generating functional introduced in ref.~\cite{Altshuler-book}.
We define the current operator $J = i/2 (\eta \otimes \partial \bar{\eta} -
\partial  \eta \otimes \bar{\eta})$. Then the generating functional for the conductance
reads
\begin{equation} \label{gfun}
 \mathcal{Z}[A] = \frac{\int \D \bar{\eta} \D \eta
    \exp \left[i \mathcal{S} + \int_r \Tr \left(J_r\,A_r\right)   \right]}{\int \D \bar{\eta} \D \eta
    \exp \left[i \mathcal{S}  \right]},
\end{equation}
where symbol $\Tr$  stands for traces over all indices.
  To average eq.(\ref{gfun}) over random potentials, we use the standard replica trick
  and introduce $N$ copies of the original system
so that the denominator in eq.(\ref{gfun}) is suppressed in the replica limit $N\to 0$.
The averaging over disorder generates in the action a quartic in $\eta$
term which can be written in the form
\begin{equation}\label{p1}
\mathcal{S}_{int} = \frac1{{4i\pi \nu_0 \tau_e}} \int_r
  ({}_i\bar{\eta}^p_{\alpha u}\ {}^j\eta^{p'}_{\beta u'})
  P^{ik}_{jl}(u,u') ({}_k\bar{\eta}^{p'}_{\beta u'}\  {}^l\eta^{p}_{\alpha u} )
\end{equation}
with the kernel
\begin{equation} \label{p2}
 P^{ik}_{jl}(u,u') =  (\tau_e / \tau_v) \delta^k_j\delta_l^i + ( \tau_e / 3\tau_s )
q_{u u'} \vec{\Sigma}^k_j \vec{\Sigma}^i_l,
\end{equation}
where we introduced the new time scale $\tau_e$ which is free at this stage as can
can seen from eqs.~(\ref{p1},\ref{p2}). In the next section we will show
that the proper normalization of the saddle point solution fixes $\tau_e$ to
the correct total mean free time.

The correlation of dimensionless  conductance can be then written as the $N\to 0$ limit
of the expression
\begin{equation}
   (\Delta G)_{uu'}^2=
\frac{1}{4\pi^2 N^4 L^4}
\left\{ \prod\limits_{v=u,u'}
   \Omega\ {\tr }\frac{\partial^2}{ \partial A_{v,r} \partial A_{v,r'} }
   \right\}
    \mathcal{Z}[A], \label{Gbarn}
\end{equation}
where $\tr$ stands only for traces  over retarded-advanced replica indices and
integration over $r$.
In order to restore the correct structure of the prefactor in front
of the Green functions product which is defined in Eq.(3)
we introduce the  matrix
 $\Omega= \sigma_0 \otimes \Upsilon_2  $ where  $\sigma_0$ and
$\Upsilon_2$ are
respectively the $2\times2$ unit matrix and matrix with all entries equal to~$1$.

Introducing a new field $Q$ of the same rank and symmetry as $\bar{\eta} \otimes \eta$
we perform the Hubbard-Stratanovich transformation on the quartic term (\ref{p1}).
The charge-conjugation symmetry ensures the invariance with respect to
the transformation $Q=C Q^* C^{-1}$ so that the hermitian matrix  $Q$  can be expressed as
\begin{eqnarray}
\label{eq:Qspin}
Q=\left(
\begin{array}{cccc}
 d_{\uparrow\uparrow} & d_{\uparrow\downarrow} & -c_{\uparrow\downarrow} & c_{\uparrow\uparrow} \\
 d_{\downarrow\uparrow} & d_{\downarrow\downarrow} & -c_{\downarrow\downarrow} & c_{\downarrow\uparrow} \\
 c^*_{\downarrow\uparrow} & c^*_{\downarrow\downarrow} & d^*_{\downarrow\downarrow} & -d^*_{\downarrow\uparrow} \\
 -c^*_{\uparrow\uparrow} & -c^*_{\uparrow\downarrow} & -d^*_{\uparrow\downarrow} & d^*_{\uparrow\uparrow}
 \end{array}
\right),
\end{eqnarray}
where the arrows denote the two electron states with the spins up and down.
This takes automatically into account the existence of two pairing channels
(i)  Diffuson modes $d$ corresponding in the standard
diagrammatics to the ladder diagrams with a small transfer momentum,
(ii)  Cooperon  modes $c$ corresponding  to the ladder diagrams with a small sum momentum
\cite{Akkermans:2007}.
 Instead of the $\uparrow,\downarrow$ basis of eq.~(\ref{eq:Qspin}), we now switch to the natural
 spin basis of the Diffusion and Cooperon Singlets and Triplets modes defined by
$d_{S}=\frac1{\sqrt{2}}(d_{\downarrow\downarrow}+d_{\uparrow\uparrow})$,
$d_{T}=\{ d_{\downarrow\uparrow},\  %
\frac1{\sqrt{2}}(d_{\downarrow\downarrow}-d_{\uparrow\uparrow}),
\ d_{\uparrow\downarrow} \}$,
$c_{S}=\frac1{\sqrt{2}}(c_{\downarrow\uparrow}-c_{\uparrow\downarrow})$, and
$c_{T}=\{ c_{\uparrow\uparrow},%
\ \frac1{\sqrt{2}}(c_{\downarrow\uparrow}+c_{\uparrow\downarrow}),
\  c_{\downarrow\downarrow} \}$.

Integrating out the bispinor fields $\eta$ we exclude all "fast" modes
and obtain the free energy  in terms of "slow" degrees of freedom $Q$
\begin{eqnarray}
  \mathcal{F}= \int_r \left\{ \frac{\pi\nu_0}{8\tau_e} Q K Q
 -\frac12 \Tr  \ln \left[ \hat{\xi} +\frac12 \left[\partial,  A \right]_{+}
+ \frac{i}{2\tau_e} Q  \right] \right\},
\label{action2-2}
\end{eqnarray}
where $K$ satisfies $P^{ik}_{jl} K_{km}^{ln} = \delta_j^n \delta^i_m$
and $[\ ,\ ]_{+}$ stands for an anticommutator.

\section{Crossover between the orthogonal and unitary universality classes}
The first step in elucidating the physics of low energy excitations is to find out the
classical solution corresponding to a spatially-homogenous field $Q$. The corresponding
saddle-point equation for $A=0$ reads
\begin{equation} \label{saddle}
\pi\nu_0 KQ =  i[\hat{\xi} + iQ/(2\tau_e)]^{-1}.
\end{equation}
Eq. (\ref{saddle}) can easily be solved in the limit of no magnetic disorder
$1/\tau_{s}=0$ which corresponds to the orthogonal universality class.
In this limit  the homogeneous solutions satisfies
the standard conditions $Q^2=\openone$ and $\Tr Q = 0$ provided that $\tau_e=\tau_v$.
One can factorize out the spin degrees  of freedom so that the saddle-point manifold
can be parameterized by $Q = U^\dagger \Lambda U $ where $U$ belongs to the coset space
$Sp(4N)/Sp(2N)\oplus Sp(2N)$ with $Sp(N)$ being the symplectic group. This insures that
$U^\dagger  U =\openone$ and that $U$ is a real quaternion matrix,
i.e. it can be expressed as $4N\times 4N$ matrix with elements being a linear combination
of quaternions with real coefficients (see {\it e.g} \cite{Efetov80}).

The presence of magnetic disorder, i.e. finite $\tau_s$, breaks the symmetry and
results in a gap for some diffusion modes.
Nevertheless, we can still enforce $Q^2=\openone$ and $\Tr Q = 0$ for
the classical solution that, however, leads to an additional
condition
\begin{equation}
KQ=Q. \label{cond1}
\end{equation}
This requirement controls  the lowering of the symmetry from the orthogonal class
to unitary one for  $q_{uu'}=1$.
Indeed, the matrix $K$ has 4 different eigenvalues out of 16 that correspond to the
Singlets and Triplets components of usual Diffuson and Cooperon modes
\begin{equation} \label{d-rates}
\lambda^{\mathrm{A}}_{\mathrm{B}} = \frac1{\tau_e}
\frac{\tau_s\tau_v}{ {\tau_s} + \kappa^{\mathrm{A}}_{\mathrm{B}}{\tau_v}},
\end{equation}
where  $\kappa^{\mathrm{D}}_{\mathrm{S}}=1$,
$\kappa^{\mathrm{D}}_{\mathrm{T}}=-1/3$, $\kappa^{\mathrm{C}}_{\mathrm{S}}=-1$,
$\kappa^{\mathrm{C}}_{\mathrm{T}}=1/3$.
The magnetic disorder cuts all those diffusion modes which eigenvalues are different
from $1$. In the unitary ensemble the only massless mode is the Diffuson Singlet so that
$\lambda^{\mathrm{D}}_{\mathrm{S}}=1$. Through this requirement we recover as expected
the Matthiessen rule  $1/\tau_e=1/\tau_v+1/\tau_s$.
Hence this rule appears in our formalism via a proper chosen normalization
of the field $Q$.

Having determined the classical solution for finite $\tau_{s}$, we can
now expand (\ref{action2-2}) around it.
The saddle-point manifold can be parameterized by
$Q = \left(1-{W}/2\right)\Lambda \left(1-{W}/2\right)^{-1}$
where $\Lambda=\textrm{diag}(\openone_{8N},-\openone_{8N})_{p,p'}$ and
$W$  is an anti-hermitian matrix anticommuting with $\Lambda$.
Performing a gradient expansion around the above homogeneous
saddle point we  rewrite the effective action as
\begin{equation}
    \mathcal{F}[W,0] = - \frac{\pi\nu_0}{8}
    \int_r \Tr \left[\frac{1}{\tau_e} W \mathcal{M} W
  + D (\partial W)^2\right]. \label{action2-3}
\end{equation}
where $D=v_F^2 \tau_e /d$ is the usual diffusion constant and we introduced the matrix
$\mathcal{M}^{il}_{jk}(u,u') = K^{il}_{jk}(u,u') - \frac12 \ \delta_k^i\
\left[  K_{nj}^{nl}(u,u) + \ K_{jn}^{ln} (u,u) \right].$
Note that the free energy (\ref{action2-2}) is invariant under the gauge transformation
$Q \to U^{\dagger}Q U$ and $A\to U^{\dagger}A U +U^{\dagger}\partial U$ \cite{Altshuler-book}.
This  gauge symmetry  ensures that the dependence of (\ref{action2-2}) on $A$ enters only
in the combination $\partial Q - [A,Q]$, and thus, this dependence
can be easily restored  in eq.~(\ref{action2-3})  .
The matrices $W$ are of the form $W=\textrm{offdiag}(B,-B^\dagger)_{p,p'}$ with $B$ satisfying the charge-conjugation symmetry $B = CB^*C\T$.
To parametrize $B$, we introduce a generalization of the
standard quaternion basis to the bispinor space \cite{Efetov80} by defining the following
basis $\phi_{4\mu+\nu} = c_{\mu \nu} \sigma_\mu \otimes \sigma_\nu$
where $c=i$ for $(0\le\mu\le2)\bigcap(1\le\nu\le3)$ and for $\mu=3,\nu=0$, and $1$ otherwise.
These matrices $\phi_\mu$ form a complete set and satisfy the charge conjugation symmetry and
relation $\Tr [\phi_\mu \phi_\nu^ \dagger ] = 4 \delta_{\mu \nu}$.
In the resulting decomposition  $B= \phi_\mu b_\mu$ and
$B^\dagger= \phi_\mu^\dagger b_\mu\T $, the $b_\mu$ are now real matrices.
Using this decomposition of matrices $B$, we obtain the following quadratic part of the
free energy   in terms of real variables $b_\mu$ :
\begin{equation}\label{eq:final}
 \mathcal{F}[b,0]  = \pi\nu_0\sum \left\{\gamma_\mu
 +  D q^2 \right\}
 b^{uu'}_{\mu \alpha\beta} b^{uu'}_{\mu \alpha\beta}.
\end{equation}
In deriving this expression, we used that  $\phi_\mu$ naturally diagonalize the
matrix $\mathcal{M}$,  and satisfy  the identity
$\phi_\mu^\dagger \mathcal{M}\phi_\nu = 4 \tau_e \delta_{\mu\nu} \gamma_\mu$.
As expected, among all masses  $\gamma_\mu$  there are  only 4 different values
\begin{equation} \label{d-rates}
\gamma^{\mathrm{A}}_{\mathrm{B}}(u,u')=
\gamma_m \frac{1- \kappa^{\mathrm{A}}_{\mathrm{B}} q_{u u'}}{1
+\kappa^{\mathrm{A}}_{\mathrm{B}} \frac{\tau_v}{\tau_s}q_{u u'}},
\end{equation}
which are the dephasing rates of Singlets and Triplets
components of Diffuson and Cooperon modes. These  rates
are computed here beyond the previous first order expansion
in $\tau_{v}/\tau_{s}$ \cite{Carpentier:2008}
and valid as long as the above classical solution is stable,
implying at least  $\tau_{v}/\tau_{s}<1$ as follows from
expression (\ref{d-rates}).
Note that
in a single disorder realization ($q_{u u'}=1$),
 the above results identify with results from diagrammatic theory even beyond the first
perturbative order in $\tau_{v}/\tau_{s}$, as can be inferred from \cite{Akkermans:2007}.
 The expressions (\ref{d-rates}) generalizes these results to
 dephasing rates between different magnetic disorder configurations
 ($q_{u,u'}\neq 1$), which are now all finite.

\section{Correlation of conductance}
\begin{figure}[tbp]
\includegraphics[clip,width=3.2 in]{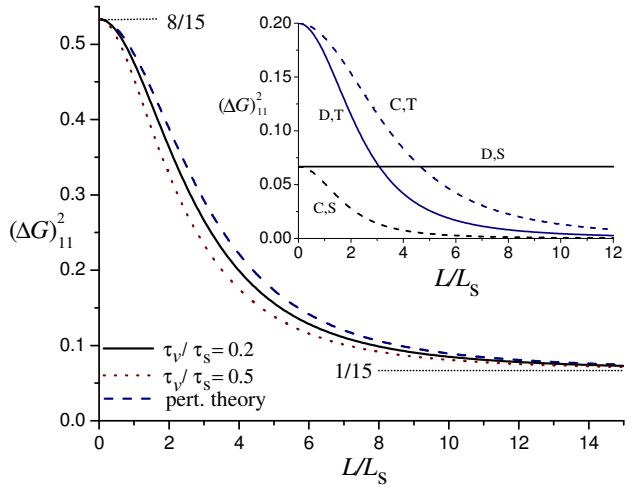}
\caption{The conductance fluctuations
$\overline{\langle  G^2 \rangle} - \overline{\langle  G\rangle}^2 $
  for diffusive wire (1D), as a function of
the wire's length $L/L_s$, and for different values of relative amplitude
$\tau_{v}/\tau_{s} $ of magnetic versus scalar  disorder.
The dashed curve corresponds to perturbative dephasing results of \cite{Carpentier:2008}. These function extrapolate
between the usual universal values 8/15 (orthogonal class) and 1/15 (unitary class).
The inset shows the contributions of the 4 different Cooperon/Diffuson modes for  $\tau_{v}/\tau_{s}=0.2 $.
}
\label{fig1}
\end{figure}
As an example of application of our formalism we derive the correlation of
conductance between two different magnetic disorder configurations.
The necessary  terms of order $A^4$ in the generating functional read
to one loop order:
$\mathcal{Z}[A]=\frac1{32}{(\pi \nu_0 D)^2}
[ \langle F_{11} F_{11}\rangle_0 + \langle F_{12} F_{12}\rangle_0 +2
       \langle F_{11} F_{12}\rangle_0]$  with vertices
$F_{11} = \int_r \Tr [{WAWA}]$ and
$F_{12} = \int_r \Tr [{WWAA}]$.
\begin{figure}[!htbp]
\includegraphics[clip,width=3.2 in]{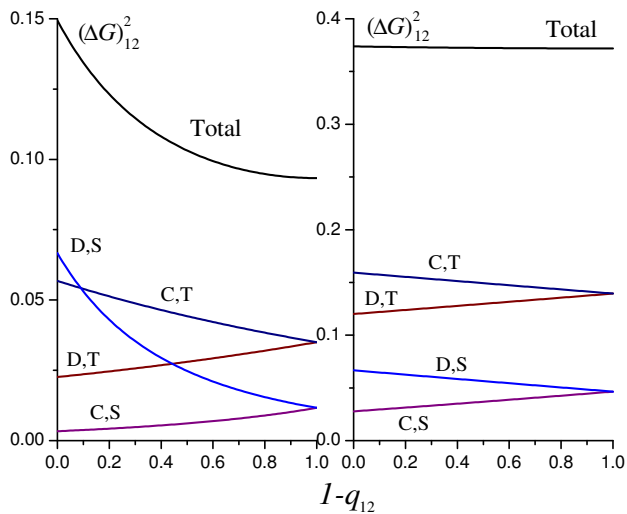}
\caption{The conductance correlations
$\overline{\langle  G(V,S_1)G(V,S_2)\rangle} - \overline{\langle   G(V,S_1)\rangle}^2 $
 as a function of the overlap $1-q$ between the different magnetic disorder (spin) configurations
 $\vec{U}_1$ and $\vec{U}_2$ for a 1D conductor.
 The relative strength of both disorders is  $\tau_v/\tau_s=0.3$, and wire length
 $L/L_s =5$ (left) and $L/L_s =1.8$ (right).
The relative contributions of Cooperon and Diffuson Singlet and Triplets modes show that
while all contribute significantly to the overall value of the correlations, their $q_{12}$
dependance is dominated by the Diffuson Singlet term.
}
\label{fig2}
\end{figure}
Expressing $W$ in the basis of $\phi_\mu$, performing Wick's contractions of $b_\mu$  and
substituting in eq.~(\ref{Gbarn})
we obtain the  correlation of conductance. The result can be conveniently
written in terms of the four dephasing lengths given by  $L^{\mathrm{A}}_{\mathrm{B}} (u,u')
=\sqrt{D/(2\gamma^{\mathrm{A}}_{\mathrm{B}})}$ and $L_s=\sqrt{D\tau_s}$ :
\begin{multline}
\label{eq:variance}
(\Delta G)_{uu'}^2 = \overline{\langle  G(V,\vec{U}_u)G(V,\vec{U}_{u'}) \rangle}
 -\overline{\langle  G(V,\vec{U}_u)\rangle}^2
\\
= f\left(\frac{L}{L^{\mathrm{D}}_{\mathrm{S}}}\right)
+3 f\left(\frac{L}{L^{\mathrm{D}}_{\mathrm{T}}}\right)
+ f\left(\frac{L}{L^{\mathrm{C}}_{\mathrm{S}}}\right)
+3 f\left(\frac{L}{L^{\mathrm{C}}_{\mathrm{T}}}\right)
\end{multline}
with $ f(x) = 3 \sum_{q\neq 0} [(Lq)^2+ x^2/2 ]^{-2}$. Specifying this result to the case $d=1$ of
a diffusive wire \cite{Pascaud98}, we use
$f(x) = 3 x^{-4} [x^2 \text{csch}^2\left(x/\sqrt{2}\right)+\sqrt{2} x
   \coth \left(x/\sqrt{2}\right)-4]$.
For $q_{uu'}=1$ the scaling function~(\ref{eq:variance}) shown for $d=1$
in fig.~\ref{fig1} gives the sample to sample conductance fluctuations and
describes the crossover between the orthogonal and unitary universality
classes. {This function extrapolates between
the magnitudes of the universal conductance
fluctuations in the systems without and with frozen magnetic impurities.}
The inset shows that all four terms contribute to the scaling function but
the correction to the unitary fixed point is dominated by Diffuson and Cooperon Triplets.
In fig.~\ref{fig2} we plot  the average correlation between conductances for different
magnetic disorder realizations, experimentally measurable through correlation of
magnetoconductances \cite{Carpentier:2008}. In the typical regime of wire length $L$
large compared to magnetic dephasing lengths of order $L_{s}$,
the correlations of conductance decay as we lower the overlap between
the corresponding spin configurations.
Moreover, this decay as a function of $1-q$ is dominated by the
dephasing of the Diffuson Singlet contribution
while other modes almost compensate for each other (see left side of fig.~\ref{fig2}).
Note that in the opposite regime $L\lesssim L_{s}$, anomalous behavior can appear with
 a $q_{12}$ dependance of $(\Delta G)_{12}^2 $ dominated by the
Diffuson Triplet and Cooperon Singlet contributions leading to a (small) increase
as a function of $1-q_{12}$.

Let us also stress that this monotonous decrease of $(\Delta G)_{uu'}^2$ in the
experimental regime of interest \cite{deVegvar:1991} $L\gtrsim L_{s}$ allows for
an interesting and
unique test of a spin glass mean-field theory.
Indeed, this theory predicts the  ultrametricity of the spin glass
phase space in the thermodynamic  limit.
According to this prediction, if we consider
three spin configurations (or three $ \vec{U}_{u}$),
and sort their mutual overlap according to $1-q_{12}\geq 1-q_{13} \geq 1-q_{23}$,
then $1-q_{12} = 1-q_{13}$.
This condition  easily translate into the practical test $(\Delta G)^2_{12} = (\Delta G)^2_{13} $ if
$(\Delta G)^2_{12} \leq (\Delta G)^2_{13} \leq (\Delta G)^2_{23} $.

{Finally, we would like to note that the
above results are valid  in the regime of coherent transport, i.e.
for samples with $L<L_{\phi}$ where $L_{\phi}=\sqrt{D/(2\gamma_{\phi})}$ is the inelastic
scattering length. This $L_{\phi}$ includes in particular contributions from possible
rare spin flips in the spin glass phase. As was shown experimentally in ref.
\cite{deVegvar:1991} one can indeed observe the
coherent electronic transport below the spin glass transition temperature $T_{SG}$
in $CuMn$ which has relatively large  $D$. This experimental observation justifies
the present study of coherent transport in the spin glass phase.}

\section{Conclusions}
We have shown how to account naturally for the magnetic dephasing of
diffusing electrons within the usual nonlinear sigma model. Motivated by the study of
mesoscopic spin glass wires, we have used this formalism to study the relative dephasing
rates between different magnetic disorder configurations, and the corresponding
correlations of conductance fluctuations amenable to direct experimental measures.
Let us finally stress that an advantage of this field theoretical method is its
flexibility, allowing for interesting extensions including the incorporation in our
approach of more complex statistical correlations of
spin configurations, along the lines of  \cite{Fisher:1986}, as well as higher
moments of the conductance correlations.

\acknowledgments  We would like to thank  F. Delduc, K. Gawedzki and
E.Orignac for numerous helpful discussions. We also acknowledge
support from the ANR through grant BLANC-06 MesoGlass.


\begin{thebibliography}{10}
\expandafter\ifx\csname url\endcsname\relax\def\url#1{\texttt{#1}}\fi

\bibitem{Houches95}
\Name{Akkermans E., Montambaux G., Pichard J.-L. \and Zinn-Justin J.} (Editors)
  \Book{Mesoscopic Quantum Physics, Les Houches Summer School} (North-Holland)
  1995.

\bibitem{Akkermans:2007}
\Name{Akkermans E. \and Montambaux G.} \Book{Mesoscopic Physics of electrons
  and photons} (Cambridge University Press) 2007.

\bibitem{Carpentier:2008}
\Name{Carpentier D. \and Orignac E.} \REVIEW{Phys. Rev. Lett.
  }{100}{2008}{057207}.

\bibitem{Vincent:2009}
\Name{Vincent E., Hammann J. \and Ocio M.} \Book{Real spin glasses relax slowly
  in the shade of hierarchical trees} presented at \Book{Wandering with
  Curiosity in Complex Landscapes} 2009.

\bibitem{Fisher:1986}
\Name{Fisher D. \and Huse D.} \REVIEW{Phys. Rev. Lett. }{56}{1986}{1601}.

\bibitem{Cugliandolo:2002}
\Name{Cugliandolo L.} \Book{Dynamics of glassy systems} presented at
  \Book{Lecture notes, Les Houches} 2002.

\bibitem{Altshuler:1985}
\Name{Al'tshuler B. \and Spivak B.} \REVIEW{JETP Lett. }{42}{1985}{447}.

\bibitem{Altshuler-book}
\Name{Altshuler B., Kravtsov V. \and Lerner I.} \Book{Mesoscopic Phenomena in
  Solids} (North-Holland, Amsterdam) 1991 p. 449.

\bibitem{Altland92}
\Name{Altland A.} \REVIEW{Z. Phys. B }{86}{1992}{101}.

\bibitem{Hikami80}
\Name{Hikami S., Larkin A. \and Nagaoka Y.} \REVIEW{Prog. Theor. Phys.
  }{63}{1980}{707}.

\bibitem{Efetov80}
\Name{Efetov K., Larkin A. \and Khmelnitskii D.} \REVIEW{Sov. Phys. JETP
  }{52}{1980}{568}.

\bibitem{Mezard:1987}
\Name{M{\'e}zard M., Parisi G. \and Virasoro M.} \Book{Spin Glass Theory and
  Beyond} (World Scientific) 1987.

\bibitem{Parisi:2007}
\Name{Parisi G.} \Book{Mean field theory of spin glasses: statics and dynamics}
  presented at \Book{Complex Systems, Les Houches Summer School}, edited by
  \Name{Bouchaud J.-P., M{\'e}zard M. \and Dalibard J.} Vol. LXXXV 2006.

\bibitem{Pascaud98}
\Name{Pascaud M. \and Montambaux G.} \REVIEW{Physics - Uspekhi}{41}{1998}{182}.

\bibitem{deVegvar:1991}
\Name{de~Vegvar P., L{\'e}vy L. \and Fulton T.}
\REVIEW{Phys. Rev. Lett.}{66}{1991}{2380}.

\end{thebibliography}
\end{document}